\documentclass[%
 preprint,
 amsmath,amssymb,
 aps,
]{revtex4-1}

\newcommand{\e}{\begin{eqnarray}}
\newcommand{\ee}{\end{eqnarray}}

\newcommand{\CN}{{\cal N}}

\def\a{\alpha}
\def\b{\beta}
\def\d{\delta}
\newcommand{\ep}{\epsilon}
\newcommand{\g}{\gamma}
\newcommand{\om}{\omega}
\newcommand{\p}{\psi}
\newcommand{\s}{\sigma}
\def\t{\tau}
\newcommand{\vp}{\varepsilon}

\newcommand{\pa}{{a^{\prime}}}
\newcommand{\pb}{{b^{\prime}}}
\newcommand{\pc}{{c^{\prime}}}
\newcommand{\pd}{{d^{\prime}}}

\newcommand{\pt}{{t^{\prime}}}
\newcommand{\pu}{{u^{\prime}}}
\newcommand{\pv}{{v^{\prime}}}
\newcommand{\pw}{{w^{\prime}}}

\begin{document}


\title{Superalgebra Realization of the 3-algebras in ${\cal N}=6, 8$ Chern-Simons-matter Theories}


\author{Fa-Min Chen}
\affiliation{Department of Physics, and State Key Laboratory of Nuclear Physics and Technology,
Peking University, Beijing 100871, China}


\date{December 4, 2011}

\begin{abstract}
We use superalgebras to realize the 3-algebras used to construct ${\cal N}=6, 8$ Chern-Simons-matter (CSM) theories. We demonstrate that the superalgebra realization of the 3-algebras provides a unified framework for classifying the gauge groups of the $\CN\geq5$ theories based on 3-algebras. Using this realization, we rederive the ordinary Lie algebra construction of the general ${\cal N}=6$ CSM theory from its 3-algebra counterpart, and reproduce all known examples as well. In particular, we explicitly construct the Nambu 3-bracket in terms of a double graded commutator of $PSU(2|2)$. The $\CN = 8$ theory of Bagger, Lambert and Gustavsson (BLG) with $SO(4)$ gauge group is constructed by using several different ways. A quantization scheme for the 3-brackets is proposed by promoting the double graded commutators as quantum mechanical double graded commutators.
\end{abstract}


\maketitle

\section{Introduction} \label{Introduction}
It was demonstrated that generic Chern-Simons gauge theories in 3D with matters
are conformally invariant at the
quantum level \cite{CSW1, CSW0, CSW2, Piguet, Saemann1}. Extended superconformal
Chern-Simons-matter (CSM) theories in 3D were conjectured to be
the dual gauge descriptions of coincident M2-branes. There are essentially two ways to construct the $\CN\geq4$ CMS theories:  the 3-algebra approach
\cite{Bagger}$-$\cite{ChenWu3}, 
and the ordinary Lie algebra approach \cite{Hosomichi:2008jb, MFM:Aug09,GaWi, ABJM, HosomichiJD}.

We begin by reviewing the Lie (2-)algebra approach. In a seminar paper \cite{GaWi}, Gaiotto and Witten (GW) have been able to construct the $\CN=4$ theory by enhancing an $\CN=1$ supersymmetry to $\CN=4$. The key point for enhancing $\CN=1$ to $\CN=4$ is to select the bosonic subalgebras of superalgebras admitting a quarternion structure as the Lie algebras of the gauge groups. The GW approach also inspired us to construct the 3-algebras in terms of superalgebras \cite{ChenWu3}. In Ref. \cite{HosomichiJD}, the $\CN=4$ GW theory was generated to include the twisted multiplets, and the $\CN=8$ BLG theories based on the Nambu 3-algebra \cite{Bagger}-\cite{Gustavsson2} was demonstrated to be equivalent to the generalized $\CN=4$ GW theory with $SU(2)\times SU(2)$ gauge group. Also, by generalizing Giaotto and Witten¡¯s idea and method, the general $\CN=5, 6$ theories were constructed rapidly \cite{Hosomichi:2008jb}. The theory of Aharony, Bergman,
Jafferis and Maldacena (ABJM) \cite{ABJM} was demonstrated to be a special case of the general $\CN=6$ theory constructed in Ref. \cite{Hosomichi:2008jb}.

We now review the 3-algebra approach. The $\CN=8$ BLG theory was first constructed by virtue of the Nambu 3-algebraic structure \cite{Bagger,Bagger2,Bagger3,Gustavsson,Gustavsson2}. Using a set of $SU(2)\times SU(2)$ $\s$-matrices to realize the Nambu 3-algebra, one can prove that the gauge symmetry generated by the Nambu 3-algebra
is $SO(4)$ \cite{Lambert1001,Bagger10, MIT}.

In an interesting paper, Bagger and Lambert (BL) have been able to
construct the general ${\cal N}=6$ theory in terms of a
hermitian 3-algebra \cite{Bagger08:3Alg}, and rederived the ${\cal N}=6,U(M)\times U(N)$ theory by using a matrix realization of the hermitian 3-algebra. While the structure constants of the Nambu 3-algebra are totally antisymmetric, the structure
constants of the hermitian 3-algebra are antisymmetric only in the
first two indices.

Using the notion of symplectic 3-algebra, another
class of ${\cal N}=6$ CSM theories, with gauge group $Sp(2M)\times
O(2)$, was constructed in Ref. \cite{ChenWu1}. One can also recast the ${\cal N}=6,U(M)\times U(N)$ into the symplectic 3-algebraic formalism \cite{ChenWu1}.
In Ref. \cite{Chen2}, we formulated the most general ${\cal N}=5,6$
CSM theories in a unified symplectic 3-algebraic framework. All examples of $\CN=5$ theories were recovered in Ref. \cite{Bagger10} by specifying the 3-algebra structure constants.

Recently, the $\CN=4,5$ theories have been constructed by using $\CN=1$ superspace formulation in a symlectic 3-algebra approach \cite{ChenWu3}. All gauge groups of the $\CN=4,5$ theories were classified by using superalgebra realization of the 3-algebras used in the $\CN=4,5$ theories \cite{ChenWu3,ChenWu4}.

In particular, our previous works demonstrate that the symplectic 3-algebra provides a unified framework for constructing all $\CN \geq 4$ theories \cite{ChenWu1, Chen2, ChenWu3}.

The symplectic 3-algebra approach must be equivalent to the Lie (2-)algebra approach, since the physical ($\CN\geq4$) theories must be the same. We find that the superalgebras are the key for proving the equivalence: On one hand, one must select the bosonic subalgebras of the superalgebras as the Lie algebras of the gauge groups of the $\CN\geq4$ theories constructed in terms of Lie algebras; on the other hand, one can use the \emph{same} superalgebras to construct the 3-algebras in the $\CN\geq4$ theories .

One goal of this paper is to use superalgebras admitting symplectic structure to realize the 3-algebras in the $\CN=6, 8$ theories, and demonstrate that the superalgebra realization of the 3-algebras provides a unified framework for classifying the gauge groups of the $\CN=5, 6, 8$ theories based on 3-algebras.

Let us first briefly review the superalgebra realization of the symplectic 3-algebra in the $\CN=4, 5$ theory. The key idea of the realization is to identify the generators of the symplectic 3-algebra $T_I$ with the fermionic generators of a superalgebra
$Q_I$, i.e., $T_I\doteq Q_I$. The 3-bracket is then naturally realized in terms of a double graded commutator: $[T_I,T_J; T_K]\doteq [\{Q_I, Q_J\}, Q_K]$. As a result, one can convert the fundamental identity (FI) of the 3-algebra into the $MMQ$ Jacobi identity of the superalgebra, with $M$ a bosonic generator. In this realization, the Lie algebra of the gauge group  generated by the 3-algebra is nothing but the bosonic part of the superalgebra, whose representation is determined by the fermionic generators. For a more mathematical approach to this subject, see Ref. \cite{MFM:Aug09, Jose, Jakob}, in which the relations between the 3-algebras and superalgebras are discussed by using representation
theory.

Since one can enhance the supersymmetry from $\CN=5$ to $\CN=6,8$ in the 3-algebraic framework \cite{Bagger10,Chen2}, one must be able to apply this realization to the $\CN=6,8$ theories. Our strategy is the following: By decomposing the general superalgebra (\ref{SLie26}) used to realize the 3-algebra in the $\CN=5$ theory, we are able to derive a general superalgebra (\ref{SLie6}) which can be used to construct the Hermitian 3-algebra in the $\CN=6$ theory \emph{and} the Nambu 3-algebra. By using the superalgebra realization of the 3-algebras, we rederive the general ${\cal N}=6$ CSM theories based on ordinary Lie algebra from its 3-algebra counterpart, and reproduce all known examples of the ${\cal N}=6$ CSM theories as well. In particular, we explicitly construct the totally antisymmetric Nambu 3-bracket in terms of a double graded commutator of the superalgebra $PSU(2|2)$, and we construct the $\CN=8$ BLG theory with $SO(4)$ gauge group by using several different ways. Since $PSU(2|2)$ also takes the form of the general superalgebra (\ref{SLie6}), it is in this sense that the superalgebra realization of the 3-algebras provides a unified framework for classifying the gauge groups of the $\CN=5, 6, 8$ theories based on 3-algebras.

Comparing with the matrix realization of 3-algebra, the superalgebra realization
has the advantage that the constraint equation imposed on the structure constants of 3-algebra for closing the Poincare superalgebra can be solved in term of the $QQQ$ Jacobi identity.

Another goal is to propose a quantization scheme for the 3-brackets, by promoting the double graded commutator as a quantum mechanical double graded commutator.

This paper is organized as follows. In Section \ref{secN4}, we review the $\CN=5,6,8$ theories based on 3-algebras. In Section \ref{Slgrlzhm3}, we realize the hermitian 3-algebra used to construct the $\CN=6$ theory by using a superalgebra, which is decomposed from the superalgebra used to realize the 3-algebra in $\CN=5$ theory. Two examples, $OSp(2|2N)$ and $U(M|N)$ are presented in this section. In Section \ref{SlgrlBLG}, we realize the Nambu 3-algebra in terms of $PSU(2|2)$, and use several different ways to construct the $\CN=8$ BLG theory. In section \ref{quantize}, a quantization scheme for the 3-brackets is proposed. Section \ref{conclusions} is devoted to conclusions. Our conventions are summarized in Appendix \ref{Identities}.
Some commutation relations of the superalgebras used in this paper are given in Appendix \ref{superalgebras}.

\section{A Review of the $\CN=5,6,8$ Theories Based on 3-Algebras }\label{secN4}
\subsection{A Short Review of Symplectic Three-Algebra }
A 3-algebra is a complex vector space equipped with a 3-bracket, mapping
three vectors to a fourth vector \cite{Chen2}:
\begin{eqnarray}\label{Symp3Bracket}
[T_I,T_J;T_K]=f_{IJK }{}^LT_{L},
\end{eqnarray}
where $f_{IJK}{}^L=f_{JIK}{}^L$ are called the structure constants of the 3-bracket.
The set of generators $T_I$ ($I=1,\cdots 2L$) are required to satisfy the fundamental
identity (FI):
\begin{equation}\label{FI}
[T_I,T_J; [T_M,T_N;T_K]]=[[T_I,T_J;T_M],T_N;
T_K]+[T_M,[T_I,T_J;T_N]; T_K]+[T_M,T_N; [T_I,T_J;T_K]].
\end{equation}
The global transformation of a field $X$ valued in this
3-algebra ($X=X^KT_K$) is defined as \cite{Bagger}:
\begin{eqnarray}\label{GlbTran}
\delta_{\tilde\Lambda}X=\Lambda^{IJ}[T_I,T_J;X],
\end{eqnarray}
where the parameter $\Lambda^{IJ}$ is hermitian. The transform (\ref{GlbTran}) is required to preserve both the
anti-symmetric form $\omega(X,Y)=\omega_{IJ}X^IY^J$, (We use $\omega_{IJ}$ and its inverse $\omega^{IJ}$ to lower and raise 3-algebra indices, respectively.) and the
Hermitian form $h(X,Y)=X^{*I}Y^I$ simultaneously
\cite{ChenWu1}:
\begin{equation}\label{sypher}
\delta_{\tilde\Lambda}\omega(X,Y)=\delta_{\tilde\Lambda}h(X,Y)=0.
\end{equation}
Imposing the reality conditions on the parameter and the antisymmetric form,
\begin{eqnarray}
\Lambda^{*IJ}=\omega_{IK}\omega_{JL}\Lambda^{LK},\quad
\omega_{IJ}^*=\omega^{IK}\omega^{JL}\omega_{KL},
\end{eqnarray}
Eqs. (\ref{sypher}) imply that the structure constants must
satisfy the reality and symmetry conditions
\begin{equation}
f_{IJKL}=f_{IJLK},\quad f^*_{IJKL}=\omega^{IM}\omega^{JN}\omega^{KO}\omega^{LP}f_{NMPO}=f^{JILK}.
\end{equation}
The 3-algebra defined by Eq. (\ref{Symp3Bracket}) $\sim$ (\ref{sypher}) is called a symplectic 3-algebra.

\subsection{A Review of the $\CN=5,6,8$ Theories Based on 3-Algebras }
\subsubsection{$\CN$=5 Theory Based on 3-Algebra}
The $\CN=5$ Lagrangian constructed by using the symplectic 3-algebra
is given by \cite{Chen2,ChenWu3}
\begin{eqnarray}\label{GeneN5Lagran}\nonumber
{\cal L}&=&\frac{1}{2}(-D_\mu\bar{Z}^A_ID^\mu
Z^I_A+i\bar{\psi}^A_I\gamma_\mu D^\mu\psi^I_A)\nonumber\\
&&-\frac{i}{2}\omega^{AB}\omega^{CD}f_{IJKL}(Z^I_AZ^K_B\psi^J_C\psi^L_D-
2Z^I_AZ^K_D\psi^J_C\psi^L_B)\nonumber\\
&&+\frac{1}{2}\epsilon^{\mu\nu\lambda}(f_{IJKL}A_\mu^{IJ}\partial_\nu
A_\lambda^{KL}+\frac{2}{3}f_{IJK}{}^Of_{OLMN}A_\mu^{IJ}A_\nu^{KL}A_\lambda^{MN})\\
&&+\frac{1}{60}(2f_{IJK}{}^Of_{OLMN}-9f_{KLI}{}^Of_{ONMJ}+2f_{IJL}{}^Of_{OKMN})Z^N_A
Z^{AI}Z^J_BZ^{BK}Z^L_CZ^{CM}.\nonumber
\end{eqnarray}
The matter fields satisfy the following reality conditions
\begin{eqnarray}\label{RealCondi}
\bar Z^{A}_{I}=\omega^{AB}\omega_{IJ}Z^J_B ,\quad
\psi^{A}_{I}=\omega^{AB}\omega_{IJ}\psi^J_B,
\end{eqnarray}
where $A$ is an $Sp(4)$ R-symmetry index, and $\omega^{AB}$ is the $Sp(4)$ invariant antisymmetric tensor.
The covariant derivative is defined as
\begin{eqnarray}
D_\mu Z^A_I &=&
\partial_\mu Z^A_I -\tilde A_\mu{}^J{}_IZ^A_J,
\end{eqnarray}
where the gauge field $\tilde A_\mu{}^J{}_I\equiv A_\mu^{KL}f_{KL}{}^J{}_I$ is anti-hermitian.
The $\CN=5$ supersymmetry transformations are given by \cite{Chen2,ChenWu3}:
\begin{eqnarray}\label{GeneSusyTransLaw}\nonumber
\delta Z^I_A&=&i\epsilon_A{}^{B\alpha}\psi^I_{B\alpha},\nonumber\\
\delta\psi^I_{A\alpha}&=&(\gamma^{\mu})_\alpha{}^{\beta}D_\mu
Z^I_B\epsilon^B{}_{A\beta}
+\frac{1}{3}f^I{}_{JKL}\omega^{BC}Z^J_BZ^K_CZ^L_D\epsilon^D{}_{A\alpha}
-\frac{2}{3}f^I{}_{JKL}\omega^{BD}Z^J_CZ^K_DZ^L_A\epsilon^C{}_{B\alpha},
\nonumber\\
\delta \tilde{A}_\mu{}^K{}_L &=&
i\epsilon^{AB\alpha}(\gamma_\mu)_\alpha{}^{\beta}\psi^J_{B\beta}Z^I_Af_{IJ}{}^K{}_L,
\end{eqnarray}
where the parameter $\epsilon^{AB}$ is antisymmetric in $AB$,
satisfying
\begin{equation}\label{SusyPara5}\nonumber
\omega_{AB}\epsilon^{AB}=0 ,\quad
\epsilon^{*}_{AB}=\omega^{AC}\omega^{BD}\epsilon_{CD}.
\end{equation}
To close the Poincare superalgebra, the structure constants must satisfy an additional constraint equation
\begin{equation}
f_{(IJK)L}=0.
\end{equation}
\subsubsection{$\CN$=6 Theory Based on 3-Algebra}\label{N6three}
The $\CN=5$ supersymmetry can be enhanced to $\CN=6$ by decomposing the $\CN=5$ fields
and the symplectic 3-algebra properly \cite{Chen2}. The reality conditions (\ref{RealCondi}) imply that the matter fields furnish a pseudo-real representation of the gauge group. If we decompose this representation as a direct sum of a complex representation and its complex conjugate representation, then the $Sp(4)$ R-symmetry can be promoted to $SU(4)$, and the $\CN=5$ supersymmetry can be enhanced to $\CN=6$ \cite{Hosomichi:2008jb}. Specifically, we decompose an $\CN=5$ scalar field into a direct summation of an $\CN=6$ scalar field $Z$ and its complex conjugate $\bar Z$ \cite{Hosomichi:2008jb}:
\begin{eqnarray}\label{DecomScaFld}
Z_A^I= Z_A^{a\alpha}=\bar{Z}^a_A\delta_{1\alpha}+
\omega_{AB}Z^B_a\delta_{2\alpha},
\end{eqnarray}
where $a=1,\cdots,L$, and $\a=1,2$. (Here $\a$ is \emph{not} an index of a spacetime
spinor. We hope this will not cause any confusion.) Similarly, we have the following decompositions for the fermionic and gauge fields:
\begin{equation}\label{fmgf}
\psi^I_A= \psi^{a\alpha}_A=\omega_{AB}\psi^{Ba}\delta_{1\alpha}
-\psi_{Aa}\delta_{2\alpha} ,\quad
\tilde{A}_\mu{}^I{}_J= \tilde{A}_\mu{}^{a\alpha}{}_{b\beta}=
\tilde{A}_\mu{}^a{}_b\delta_{1\alpha}\delta_{1\beta}
-\tilde{A}_\mu{}^b{}_a\delta_{2\alpha}\delta_{2\beta},
\end{equation}
where $\tilde{A}_\mu{}^a{}_b\equiv A_{\mu}{}^d{}_cf^{ac}{}_{db}$ is an anti-hermitian $\CN=6$ gauge field, with $f^{ac}{}_{db}$ the structure constants of the hermitian 3-algebra. The $\CN=6$ fields satisfy the reality conditions
\begin{equation}\label{N6RealCondi}
Z^{*A}_a= \bar Z_A^{a},\quad \psi^{*Aa}=\psi_{Aa}, \quad \tilde{A}^*_\mu{}^a{}_b=- \tilde{A}_\mu{}^b{}_a.
\end{equation}
To be consistent with the reality conditions of the $\CN=6$ matter fields, we must decompose the antisymmetric tensor $\omega_{IJ}$  as
\begin{eqnarray}\label{DecomMetr}
\omega_{IJ}=
\omega_{a\alpha,b\beta}=\delta_a{}^b\delta_{1\alpha}\delta_{2\beta}
-\delta^a{}_b\delta_{2\alpha}\delta_{1\beta}.
\end{eqnarray}
The decomposition of the gauge field (see the second equation of (\ref{fmgf}))
follows from the decomposition of the structure constants given by
\begin{eqnarray}\label{DecomStrucConst}\nonumber
f_{IJKL}= f_{a\alpha,b\beta,c\gamma,d\delta}
&=&f^{ac}{}_{db}\delta_{2\alpha}\delta_{1\beta}\delta_{2\gamma}\delta_{1\delta}
+f^{ad}{}_{cb}\delta_{2\alpha}\delta_{1\beta}\delta_{1\gamma}\delta_{2\delta}\\&&+f^{bc}{}_{da}\delta_{1\alpha}\delta_{2\beta}\delta_{2\gamma}\delta_{1\delta}
+f^{bd}{}_{ca}\delta_{1\alpha}\delta_{2\beta}\delta_{1\gamma}\delta_{2\delta},
\end{eqnarray}
combined with the decomposition of $A_\mu^{IJ}$ given by
\begin{eqnarray}\label{DecomGauFld}
A_\mu^{IJ}=
A_\mu^{a\alpha,b\beta}=-\frac{1}{2}(A_\mu{}^a{}_b\delta_{1\alpha}\delta_{2\beta}
+A_\mu{}^b{}_a\delta_{2\alpha}\delta_{1\beta}) .
\end{eqnarray}
The generator $T_I$ is decomposed into
\begin{eqnarray}\label{DecomT}
T_{I}=T_{a\alpha} &=&\bar
t_{a}\delta_{1\alpha}-t^{a}\delta_{2\alpha},
\end{eqnarray}
where $t^a$ is a generator of the ($\CN=6$) hermitian 3-algebra, and
$\bar{t}_a\equiv t^{*a}$ is its complex conjugate.
With these decompositions, the fundamental identity (\ref{FI})
reduces to
\begin{equation}\label{N6FI}
f^{fc}{}_{dg}f^{ag}{}_{eb}-f^{af}{}_{gb}f^{gc}{}_{de}
+f^{cf}{}_{eg}f^{ag}{}_{db}-f^{ac}{}_{gb}f^{gf}{}_{ed}=0.
\end{equation}
Here the 3-bracket of the ($\CN=6$)  hermitian 3-algebra is defined as
\begin{eqnarray}\label{N6Bracket}
[t^a,t^c;\bar{t}_b]=f^{ac}{}_{bd}t^d .
\end{eqnarray}
Also the constraint condition $f_{(IJK)L}=0$ and the reality condition imposed on the structure constants reduce to
\begin{eqnarray}\label{SymmeOfN6F}
f^{ab}{}_{cd}=-f^{ba}{}_{cd},\quad f^{*ab}{}_{cd}=f^{cd}{}_{ab}.
\end{eqnarray}
Now Eq. (\ref{DecomStrucConst}) is equivalent to
\begin{eqnarray}\label{DecomBracket}\nonumber
[T_I,T_J; T_K]&= &[T_{a\alpha},T_{b\beta};
T_{c\gamma}]\\
\nonumber&=&[t^a,t^c;\bar{t}_b]\delta_{2\alpha}\delta_{1\beta}\delta_{2\gamma}
+[t^a,t^c;\bar{t}_b]^*\delta_{1\alpha}\delta_{2\beta}\delta_{1\gamma}\\
&&
+[t^b,t^c;\bar{t}_a]\delta_{1\alpha}\delta_{2\beta}\delta_{2\gamma}
+[t^b,t^c;\bar{t}_a]^*\delta_{2\alpha}\delta_{1\beta}\delta_{1\gamma}.
\end{eqnarray}
The hermitian 3-algebra defined by Eqs. (\ref{N6FI}) $\sim$ (\ref{SymmeOfN6F}) is nothing but the 3-algebra used by BL to construct the general $\CN=6$ theory \cite{Bagger08:3Alg}.

Substituting the decompositions of the fields and the structure constants into the $\CN=5$ lagrangian (\ref{GeneN5Lagran}) reproduces the ${\cal N}=6$ Lagrangian in Ref. \cite{Bagger08:3Alg}:
\begin{eqnarray}\label{N6Lagrangian}
\nonumber {\cal L} &=& -D_\mu
\bar{Z}_A^aD^\mu Z^A_a - i\bar\psi^{Aa}\gamma^\mu D_\mu\psi_{Aa}\\
\nonumber && -if^{ab}{}_{cd}\bar\psi^{Ad} \psi_{Aa}
Z^B_b\bar{Z}_B^c+2if^{ab}{}_{cd}\bar\psi^{Ad}
\psi_{Ba}Z^B_b\bar{Z}_A^c\\ \nonumber
&&-\frac{i}{2}\varepsilon_{ABCD}f^{ab}{}_{cd}\bar\psi^{Ac}
\psi^{Bd}Z^C_aZ^D_b -\frac{i}{2}\varepsilon^{ABCD}f^{cd}{}_{ab}
\bar\psi_{Ac}\psi_{Bd}\bar{Z}_C^a\bar{Z}_D^b \\
&&+\frac{1}{2}\varepsilon^{\mu\nu\lambda}
(f^{ab}{}_{cd}A_{\mu}{}^c{}_b\partial_\nu A_{\lambda}{}^d{}_a
+\frac{2}{3}f^{ac}{}_{dg}f^{ge}{}_{fb}
A_{\mu}{}^b{}_aA_{\nu}{}^d{}_c A_{\lambda}{}^f{}_e)\\ \nonumber &&
-\frac{2}{3}(f^{ab}{}_{cd}f^{ed}{}_{fg}
-\frac{1}{2}f^{eb}{}_{cd}f^{ad}{}_{fg})\bar{Z}_A^c Z^A_e\bar{Z}_B^f
Z^B_a\bar{Z}_D^g Z^D_b.
\end{eqnarray}
And the  ${\cal N}=6$ SUSY transformation law reads
\begin{eqnarray}\label{N6susy}
\nonumber  \delta Z^A_d &=& -i\bar\epsilon^{AB}\psi_{Bd} \\
 \nonumber
\delta \psi_{Bd} &=& \gamma^\mu D_\mu Z^A_d\epsilon_{AB} +
  f^{ab}{}_{cd}Z^C_aZ^A_b \bar{Z}_{C}^{c} \epsilon_{AB}+f^{ab}{}_{cd}
  Z^C_a Z^D_{b} \bar{Z}_{B}^{c}\epsilon_{CD} \\
 \delta \tilde A_\mu{}^c{}_d &=&
-i\bar\epsilon_{AB}\gamma_\mu Z^A_a\psi^{Bb} f^{ca}{}_{bd} +
i\bar\epsilon^{AB}\gamma_\mu \bar{Z}_{A}^{a}\psi_{Bb}f^{cb}{}_{ad}.
\end{eqnarray}
Here the SUSY transformation parameters $\epsilon_{AB}$ satisfy
\begin{equation}
\epsilon_{AB}=-\epsilon_{BA},\quad
\epsilon^*_{AB}=\epsilon^{AB}
=\frac{1}{2}\varepsilon^{ABCD}\epsilon_{CD}.
\end{equation}
\subsubsection{${\cal N}$=8 Theory Based on 3-Algebra} \label{secBLG}
If the inner product becomes the standard inner
product in the Euclidian space
\begin{eqnarray}\label{EucInnerProd}
h(X,Y)=X_aY_a \quad {\rm or}\quad h(t^a,t^b)=\delta^{ab},
\end{eqnarray}
then there is no difference between a lower index $a$ and an upper
index $a$, i.e., $\bar{t}_a=\bar t^a$. As a result, the 3-bracket
(\ref{N6Bracket}) becomes
\begin{eqnarray}\label{NambuBracket}
[t^a,t^c,\bar t^b]=f^{acb}{}_{d}t^d.
\end{eqnarray}
If the first 3 indices of $f^{acb}{}_{d}$ are antisymmetric, then
Eq. (\ref{NambuBracket}) becomes the famous Nambu 3-bracket; and Eqs.
(\ref{SymmeOfN6F}) imply that
\begin{equation}\label{nambustr}
f^{abcd}\equiv \delta^{de}f^{abc}{}_e
\end{equation}
are \emph{totally antisymmetric}. Now the FI (\ref{N6FI}) can be
converted into
\begin{eqnarray}\label{BLGFI}
f^{afe}{}_{g}f^{cdg}{}_{b}-f^{cda}{}_{g}f^{gfe}{}_{b}-f^{cdf}{}_{g}f^{age}{}_{b}-f^{cde}{}_{g}f^{afg}{}_{b}=0.
\end{eqnarray}
The 3-algebra defined by Eq. (\ref{NambuBracket}) $\sim$
(\ref{BLGFI}) is nothing but the Nambu 3-algebra. It was proved that the only non-trivial possibility is that $f^{abcd}=\varepsilon^{abcd}$ (up to a constant). Here $\varepsilon^{abcd}$ is the familiar Levi-Civita tensor. And the gauge group generated by the Nambu 3-algebra is $SO(4)$ \cite{Gauntlett, Papadopoulos, MIT}. In Ref. \cite{Bagger10}, it has been demonstrated explicitly that the $\CN=6$ supersymmetry is promoted to $\CN=8$ if the structure constants $f^{abcd}\propto\varepsilon^{abcd}$. So substituting Eq.
(\ref{nambustr}) into (\ref{N6Lagrangian}) gives
the $\CN=8$ BLG theory.

\section{$\CN$=6, 8 Theories in Terms of the Bosonic parts of Superalgebras}\label{LieN6N8}
In this section, we first try to find a superalgebra which can
be used to realize the hermitian 3-algebra and the Nambu 3-algebra.
We then derive the ordinary Lie algebra constructions of the $\CN=6,
8$ theories by using the superalgebra realization of
3-algebras.

\subsection{$\CN=6$ Theories in Terms of the Bosonic Parts of Superalgebras}\label{Slgrlzhm3}
\subsubsection{General $\CN=6$ Theory}
In this section, we determine the superalgebra which can be used to realize the
hermitian 3-algebra and the Nambu 3-algebra, and classify the gauge groups of the
$\CN=6$ theory. Recall that we used the superalgebra
\e\label{SLie26} &&[M^m, M^n]=C^{mn}{}_pM^p,\nonumber\\
&&[M^m, Q_I]=-\t^m_{IJ}\omega^{JK}Q_K,\nonumber\\
&&\{Q_I,Q_J\}=\t^m_{IJ}k_{mn}M^n.\ee
to realize the symplectic 3-algebra \cite{ChenWu3}. Here $\omega^{JK}$ is an invariant antisymmetric tensor, and $k_{mn}$ an invariant quadratic form. We connected the symplectic 3-algebra with the above superalgebra by identifying the 3-algebra generators $T_I$ with the fermionic generators $Q_I$:
\begin{equation}\label{TQE}
T_I\doteq Q_I,
\end{equation}
and by presenting the 3-bracket as a double graded bracket \cite{ChenWu3}, i.e.,
\begin{equation}\label{3bredb}
[T_I,T_J; T_K]\doteq [\{Q_I, Q_J\}, Q_K]=k_{mn}\t^m_{IJ}\t^n_K{}^LQ_L.
\end{equation}
In this realization, the structure constants of the 3-algebra are just the structure constants of the double graded commutator, i.e.,
\e\label{tensorpr}
f_{IJKL}=k_{mn}\t^m_{IJ}\t^n_{KL}.
\ee
The right hand side of (\ref{tensorpr}) was first introduced in the $\CN=4$ GW theory to construct the superspace coupling \cite{GaWi}; and its property $k_{mn}\t^m_{(IJ}\t^n_{K)L}=0$, implied by the $Q_IQ_JQ_K$ Jacobi identity, strongly hints that one can construct the 3-bracket in terms of the double graded commutator (\ref{3bredb}) \cite{ChenWu3}.
With Eq. (\ref{TQE}),
Eq. (\ref{DecomT}) becomes
\begin{eqnarray}\label{DecomQ}
T_I\doteq Q_I= Q_{a\alpha}
=\bar{Q}_{a}\delta_{1\alpha}-Q^{a}\delta_{2\alpha}.
\end{eqnarray}
Recall that $Q_I$ furnish a pseudo-real (quaternion) representation
of the bosonic part of the superalgebra (\ref{SLie26}) \cite{ChenWu3}, and we
decompose this pseudo-real representation into a complex
representation and its complex conjugate representation for
promoting the $\CN=5$ supersymmetry to $\CN=6$. So, with the
decomposition (\ref{DecomQ}), if the fermionic generators $Q^{a}$
furnish a complex representation of the bosonic part of
(\ref{SLie26}), then $\bar{Q}_a$ must furnish a complex conjugate
representation of the bosonic part of (\ref{SLie26}). Namely,
we must have
\begin{equation}
[M^m, Q^a]=-\t^{ma}{}_bQ^b\quad {\rm and}\quad [M^m, \bar
Q_a]=\t^{mb}{}_a\bar Q_b,
\end{equation}
where $\t^{ma}{}_b$ is anti-hermitian, i.e.,
\begin{equation}\label{rltcd}
\t^{*ma}{}_b=-\t^{mb}{}_a.
\end{equation}
Substituting
\begin{equation}\label{DcmQp}
Q_{I}=\bar Q_{a}\delta_{1\alpha}-Q^{a}\delta_{2\alpha}
\end{equation}
into the LHS of the second equation of (\ref{SLie26}) gives
\begin{equation}
[M^m, \bar Q_a\d_{1\a}-Q^a\d_{2\a}]=\t^{mb}{}_a\bar
Q_b\d_{1\a}+\t^{ma}{}_bQ^b_{2\a}.
\end{equation}
Comparing the RHS of the above equation with the RHS of the second equation of
(\ref{SLie26}), we obtain
\begin{equation}\label{RRStar}
\t^{mJ}{}_I=\t^{mb}{}_a\d_{1\a}\d_{1\b}-\t^{ma}{}_b\d_{2\a}\d_{2\b}.
\end{equation}
By (\ref{rltcd}), the RHS is a direct sum of $\t^{mb}{}_a$ and its
complex conjugate. So the pseudo-real representation is indeed
decomposed into a complex representation and its complex conjugate
representation. Substituting (\ref{DcmQp}) and (\ref{RRStar}) into
the LHS and RHS of the third equation of (\ref{SLie26}),
respectively, we obtain
\begin{equation}\label{DcmQSlie}
\{\bar Q_{a}\delta_{1\alpha}-Q^{a}\delta_{2\alpha}, \bar
Q_{b}\delta_{1\b}-Q^{b}\delta_{2\b}\}
=-(\t^{mb}{}_ak_{mn}M^n\d_{1\a}\d_{2\b}+\t^{ma}{}_bk_{mn}M^n\d_{2\a}\d_{1\b}),
\end{equation}
where we have used Eq. (\ref{DecomMetr}). The anticommutators can
be easily read off from the above equation:
\begin{equation}\label{anticmtt}
\{Q^b, \bar Q_{a}\}=\t^{mb}{}_ak_{mn}M^n,\quad \{\bar Q_{a}, \bar
Q_b\}=\{Q^{a}, Q^b\}=0.
\end{equation}
In summary, the superalgebra used to realize the $\CN=6$
hermitian 3-algebra is the following:
\e\label{SLie6} &&[M^m, M^n]=C^{mn}{}_sM^s,\nonumber\\
&&[M^m, Q^a]=-\t^{ma}{}_bQ^b,\quad [M^m, \bar
Q_a]=\t^{mb}{}_a\bar Q_b,\nonumber\\
&&\{Q^a, \bar Q_{b}\}=\t^{ma}{}_bk_{mn}M^n,\quad \{\bar Q_{a}, \bar
Q_b\}=\{Q^{a}, Q^b\}=0.\ee
In this way, we rederive the above superalgebra by decomposing
the superalgebra (\ref{SLie26}) properly. The superalgebra used to realize construct the Nambu 3-algebra must also take the form of (\ref{SLie6}), since the Nambu 3-algebra is a special case of the $\CN=6$ hermitian 3-algebra. The superalgebras $OSp(2|2N)$ and $U(M|N)$ (or its cousins $SU(M|N)$ and $PSU(M|N)$) take the form of (\ref{SLie6}); their commutation relations are given in Appendix \ref{superalgebras}. Later we will see that (\ref{SLie6}) indeed can be used to realize the 3-algebras in the $\CN=6, 8$ theory. Namely, the superalgebra realization of the 3-algebras can provides a unified framework for classifying the gauge groups of the $\CN=5, 6, 8$ theories based on 3-algebras, since (\ref{SLie6}) is decomposed from (\ref{SLie26}).

With these decompositions, the double graded commutator
\begin{equation}\label{dcommutator26}
[\{Q_I,Q_J\}, Q_K]=k_{mn}\tau^m_{IJ}\tau^n_{K}{}^LQ_L
\end{equation}
is decomposed into two sets:
\begin{equation}\label{dcmmtt6}
[\{Q^b,\bar Q_a\}, Q^c]=-k_{mn}\tau^{mb}{}_a\tau^{nc}{}_dQ^d,\quad
[\{Q^a,\bar Q_b\}, \bar Q_c]=k_{mn}\tau^{ma}{}_b\tau^{nd}{}_c\bar
Q_d.
\end{equation}
However, their structure constants are related by a reality
condition (see Eqs. (\ref{sltN6str}) and (\ref{RealCondiOnN6FN})).
So we need only to consider the first equation. Using Eqs. (\ref{3bredb}) and (\ref{DecomQ}), and comparing the
decomposition of (\ref{dcommutator26}) with (\ref{DecomBracket}), we
are led to the following equations:
\begin{eqnarray}\label{N6brckrlzz}
&&[t^b, t^c; \bar t_a]\doteq[\{Q^b,\bar Q_a\},
Q^c]=-k_{mn}\tau^{mb}{}_a\tau^{nc}{}_dQ^d,\\
&&[t^b,t^c; \bar t_a]^{\ast}\doteq-[\{Q^a,\bar Q_b\}, \bar
Q_c]=-k_{mn}\tau^{ma}{}_b\tau^{nd}{}_c\bar Q_d.\label{n6strc}
\end{eqnarray}
where the LHS of the first equation is the 3-bracket of the
hermitian 3-algebra, and $t^a$ are the generators of the hermitian
3-algebra (see section \ref{N6three}). Here we have used the superalgebra realization of the hermitian 3-algebra:
\begin{equation}
t^{a}\doteq Q^a,\quad \bar t_{a}\doteq \bar
Q_{a}.
\end{equation}
The structure constants can be
read off immediately from (\ref{n6strc}):
\begin{equation}\label{sltN6str}
f^{bc}{}_{ad}=-k_{mn}\t^{mb}{}_a\t^{nc}{}_d.
\end{equation}
It is straightforward to verify that the above tensor product is a
solution of the FI (\ref{N6FI}) of the hermitian 3-algebra (for
convenience, we cite it here):
\begin{eqnarray}\label{N6FI4Q}
f^{fc}{}_{dg}f^{ag}{}_{eb}-f^{af}{}_{gb}f^{gc}{}_{de}
+f^{cf}{}_{eg}f^{ag}{}_{db}-f^{ac}{}_{gb}f^{gf}{}_{ed}=0.
\end{eqnarray}
The solution (\ref{sltN6str}) was first discovered by BL
\cite{Bagger08:3Alg}, using a different approach. Similarly, the
$Q_IQ_JQ_K$ Jacobi identity is decomposed into two sets: the
$Q^bQ^c\bar Q_a$ Jacobi identity and the $\bar Q_b\bar Q_cQ^a$
Jacobi identity. Let us examine the $Q^bQ^c\bar Q_a$ Jacobi
identity:
\begin{equation}\label{QQbQ}
[\{Q^b,\bar Q_a\}, Q^c]+[\{\bar Q_a\, Q^c\}, Q^b]+[\{Q^c, Q^b\},
\bar Q_a]=0.
\end{equation}
By $\{Q^c, Q^b\}=0$, the last term of the LHS vanishes. The equation
for the remaining two terms implies that
\begin{equation}
k_{mn}\t^{mb}{}_a\t^{nc}{}_d+k_{mn}\t^{mc}{}_a\t^{nb}{}_d=0.
\end{equation}
Namely, the structure constants $f^{bc}{}_{ad}$ are antisymmetric in
the first two indices :
\begin{eqnarray}\label{SymmeOfN6FN}
f^{bc}{}_{ad}=-f^{cb}{}_{ad}.
\end{eqnarray}
Also, the reality condition $(\ref{rltcd})$ implies that the
structure constants satisfy the reality condition:
\begin{eqnarray}\label{RealCondiOnN6FN}
f^{*ab}{}_{cd}=f^{cd}{}_{ab}.
\end{eqnarray}
Eqs (\ref{SymmeOfN6FN}) and (\ref{RealCondiOnN6FN}) are nothing but the two equations in
(\ref{SymmeOfN6F}).

Here we would like to demonstrate that the FI (\ref{N6FI4Q}),
satisfied by the structure constants, is equivalent to the $MMQ$ or
$MM\bar Q$ Jacobi identity of the superalgebra (\ref{SLie6}).
Substituting (\ref{DecomQ}) into the FI (\ref{FI}), the later is decomposed into eight
sets; one of them reads
\begin{eqnarray}\label{FIHM}
&&[\{\bar Q_a,Q^b\},[\{\bar Q_e,Q^f\},Q^c]]\\&=&[\{[\{\bar
Q_a,Q^b\}\bar Q_e],Q^f\}, Q^c]+[\{\bar Q_e,[\{\bar Q_a,Q^b\},Q^f]\},
Q^c]+[\{\bar Q_e,Q^f\},[\{\bar Q_a,Q^b\},Q^c]].\nonumber
\end{eqnarray}
Substituting (\ref{dcmmtt6}) into this equation shows that it
precisely coincides with the FI (\ref{N6FI4Q}). The rest (seven)
sets can be also converted into the FI (\ref{N6FI4Q}). So it is
sufficient to examine Eq. (\ref{FIHM}). On the other hand, by using
the superalgebra (\ref{SLie6}), one can convert Eq.
(\ref{FIHM}) into the following equation:
\e \t^{mb}{}_{a}\t^{nf}{}_{e}([M_n,[M_m,Q^c]]-[M_m,[M_n,Q^c]]+[[M_m,
M_n],Q^c])=0,\ee
which is the $MMQ$ Jacobi identity of the superalgebra
(\ref{SLie6}). 
With $Q^c$ replaced by $\bar Q_c$, Eq.
(\ref{FIHM}) becomes another set FI decomposed from (\ref{FI}). It
can be converted into $MM\bar Q$ Jacobi identity of (\ref{SLie6}).
Therefore, the FI (\ref{N6FI4Q}) is indeed equivalent to the $MMQ$
or $MM\bar Q$ Jacobi identity of the superalgebra
(\ref{SLie6}).

Now consider the basic definition of the transformation
\begin{equation}
\d_{\tilde\Lambda}X^a=\Lambda^c{}_df^{ad}{}_{cb}X^b.
\end{equation}
Substituting (\ref{sltN6str}) into the above equation gives
\begin{equation}\label{lietran}
\d_{\tilde\Lambda}X^a=\Lambda^c{}_dk_{mn}\t^{md}{}_{c}\t^{na}{}_{b}X^b.
\end{equation}
From the Lie group point of view, this is a gauge transformation by
a parameter $\hat\Lambda^m\equiv\Lambda^c{}_d\t^{md}{}_{c}$. On the other hand, by the $MMQ$ Jacobi identity of (\ref{SLie6}), we learn that $\t^{na}{}_{b}$ is a matrix
representation of the bosonic generator $M^n$, i.e.,
 \begin{equation}
[\t^m,\t^n]^a{}_b=C^{mn}{}_p\t^{pa}{}_b.
\end{equation}
 This matrix representation is furnished by the fermionic generators of (\ref{SLie6}) (see the second line of (\ref{SLie6})). Therefore, the Lie algebra of the gauge group generated by the 3-algebra is just the bosonic subalgebra of (\ref{SLie6}). And the representation of the matter
fields is determined by the fermionic generators of (\ref{SLie6}). The quadratic form $k_{mn}$ plays a fundamental role in our construction, since it appears both in (\ref{sltN6str}) and (\ref{lietran}).

With the solution for the structure constants of the hermitian 3-algebra
\begin{equation}\label{soln}
f^{ab}{}_{cd}=k_{mn}\t^{ma}{}_{d}\t^{nb}{}_{c},\quad [\t^m,
\t^n]^a{}_b=C^{mn}{}_p\t^{pa}{}_b,
\end{equation}
we can rewrite the gauge field as
\begin{equation}\label{gauge}
\tilde
A_\mu{}^a{}_b=A_\mu{}^{d}{}_cf^{ac}{}_{db}=A_\mu{}^{d}{}_ck_{mn}\t^{na}{}_{b}\t^{mc}{}_{d}\equiv
\hat A^m_\mu k_{mn}\t^{na}{}_{b}.
\end{equation}
Following Ref. \cite{GaWi}, we define the $\CN=6$ `momentum map' and
`current' operators as
\e \nonumber\mu^{mA}{}_B&\equiv& \t^{ma}{}_bZ_a^A\bar{Z}^b_B, \quad \mu^{mnA}{}_B\equiv\{\t^m,\t^n\}^a{}_bZ_a^A\bar{Z}^b_B, \\
j^{mAB}&\equiv& \t^{ma}{}_bZ_a^A\p^{bB},\quad \bar{j}^{m}_{AB}\equiv \t^{mb}{}_a\bar{Z}^a_A\p_{bB}.
\ee
Substituting the (\ref{soln}) and (\ref{gauge}) into the Lagrangian
(\ref{N6Lagrangian}) and the SUSY law (\ref{N6susy}) gives the
ordinary Lie algebra construction of the $\CN=6$ Theory. Here the $\CN=6$ Lagrangian reads
\begin{eqnarray}\label{N6LagrangianLie}
\nonumber {\cal L} &=& -D_\mu
\bar{Z}_A^aD^\mu Z^A_a - i\bar\psi^{Aa}\gamma^\mu D_\mu\psi_{Aa}\\
\nonumber && -\frac{i}{2}k_{mn}(-2j^{mAB}\bar{j}^n_{AB}+4j^{mAB}\bar{j}^n_{BA}
+\varepsilon_{ABCD}j^{mAB}j^{nCD}+\varepsilon^{ABCD}\bar{j}^m_{AB}\bar{j}^n_{CD}) \\
&&+\frac{1}{2}\varepsilon^{\mu\nu\lambda}
(k_{mn}\hat A_{\mu}^m\partial_\nu \hat A_{\lambda}^n
+\frac{1}{3}\tilde C_{mnp}
\hat A_{\mu}^m \hat A_{\nu}^n \hat A_{\lambda}^p)\\ \nonumber &&
+\frac{1}{6}\tilde C_{mnp}\mu^{mA}{}_B\mu^{nB}{}_{C}\mu^{pC}{}_A
+\frac{1}{2}\mu^{mnA}{}_B\tilde\mu_m^B{}_C\tilde\mu_n^C{}_A,
\end{eqnarray}
where
\e\label{tdC}
\tilde C_{mnp}\equiv k_{ms}k_{nq}C^{sq}{}_p,\quad\tilde \mu_m^B{}_C\equiv k_{mn}\mu^{nB}{}_C.
\ee
In deriving the Lagrangian, we have used the identity $-\varepsilon^{ABCD}
=\omega^{AB}\omega^{CD}-\omega^{AC}\omega^{BD}+\omega^{AD}\omega^{BC}$, whose LHS differs a minus sign from the one in Ref. \cite{Hosomichi:2008jb}.
The  ${\cal N}=6$ SUSY transformations are given by
\begin{eqnarray}\label{N6susyLie}
\nonumber  \delta Z^A_d &=& -i\bar\epsilon^{AB}\psi_{Bd} \\
 \nonumber
\delta \psi_{Bd} &=& \gamma^\mu D_\mu Z^A_d\epsilon_{AB} +
  \tilde\mu_m^A{}_C\t^{ma}{}_dZ^C_a\epsilon_{AB}+\tilde\mu^D_m{}_B\t^{ma}{}_d
  Z^C_a \epsilon_{CD} \\
 \delta \hat A^m_\mu{}&=&
-i\bar\epsilon_{AB}\gamma_\mu j^{mAB}+
i\bar\epsilon^{AB}\gamma_\mu \bar{j}^m_{AB}.
\end{eqnarray}
The parameters $\epsilon_{AB}$ satisfy
\begin{equation}
\epsilon_{AB}=-\epsilon_{BA},\quad
\epsilon^*_{AB}=\epsilon^{AB}
=\frac{1}{2}\varepsilon^{ABCD}\epsilon_{CD}.
\end{equation}
The Lagrangian (\ref{N6LagrangianLie}) and the supersymmetry transformation law (\ref{N6susyLie}) are in agreement with the ones constructed directly in an ordinary Lie algebra approach \cite{Hosomichi:2008jb}.

The
bosonic parts of the superalgebras $OSp(2|2N)$ and $U(M|N)$ (or
its cousins $SU(M|N)$ and $PSU(M|N)$) can be selected as the Lie
algebras of the gauge groups of the $\CN=6$ Theory (see
(\ref{SLie6})). In particular, if the superalgebra is
$PSU(2|2)$, then (\ref{N6brckrlzz}) becomes the Nambu bracket, and
the $\CN=6$ supersymmetry gets enhanced to $\CN=8$. This will be the
topic of section \ref{SlgrlBLG}.
\subsubsection{$OSp(2|2N)$}
If we specify the superalgebra (\ref{SLie6}) as $OSp(2|2N)$ (the commutation relations of $OSp(2|2N)$ are given by (\ref{sp22n2})), the double graded
commutator (\ref{N6brckrlzz}) becomes
\begin{eqnarray}\label{SyplStruc1}
[\{Q_{a-},Q_{c+}\},Q_{b-}]&=&f_{a-,b-,c+,d+}Q^{d+}\\\nonumber &=& k[(\omega_{ab}\omega_{cd}
-\omega_{ad}\omega_{bc})h_{-+}h_{-+}
-(\omega_{ac}\epsilon_{-+})(\omega_{bd}\epsilon_{-+})]Q^{d+}.
\end{eqnarray}
(In the
Lagrangian (\ref{N6Lagrangian}), the index $a$
runs from 1 to $L$. Here we split $a$ into two
indices: $a\rightarrow a\pm$, by setting $L=4N$. We hope this will not
cause any confusion.) Here $a,b=1,2,\cdots,2N$ are the $Sp(2N)$
indices while $+,-$ the $SO(2)$ indices.  We use the gauge invariant
antisymmetric tensor $\omega^{a+,b-}\equiv \omega^{ab}h^{+-}$ to
raise the first two pairs of indices of the structure constants in
(\ref{SyplStruc1}):
\begin{eqnarray}
f^{a+b+}{}_{c+d+} = -k[(\omega^{ab}\omega_{cd}
+\delta^a{}_d\delta^b{}_c)\delta^{+}{}_{+} \delta^{+}{}_{+} +
(\delta^a{}_c)(-i
\delta^{+}{}_{+})(\delta^b{}_d)(-i\delta^{+}{}_{+})].
\end{eqnarray}
Suppressing the $SO(2)$ indices gives
\begin{eqnarray}\label{SyplStruc}
f^{ab}{}_{cd} &=&-k(\omega^{ab}\omega_{cd} +\delta^a{}_d\delta^b{}_c
-\delta^a{}_c{}\delta^b{}_d),
\end{eqnarray}
which are precisely the same as that of our previous paper \cite{ChenWu1} up to an unimportant overall sign. The above specified structure constants have the correct symmetry properties and satisfy the reality condition. Substituting
(\ref{SyplStruc}) into (\ref{N6Lagrangian}) and (\ref{N6susy}) give the $\CN=6, Sp(2N)\times U(1)$ theory \cite{ChenWu1}. This theory can be derived in an alternative way \cite{Hosomichi:2008jb}: one can first read off the representation $\t^{ma+}{}_{b+}$, the invariant quadratic form $k_{mn}$ and the structure constants $C^{mn}{}_p$ from the superalgebra (\ref{sp22n2}), then plug them into the Lagrangian (\ref{N6LagrangianLie}) and the SUSY transformations (\ref{N6susyLie}).
\subsubsection{$U(M|N)$ }
If we specify the superalgebra (\ref{SLie6}) as $U(M|N)$ (the commutation relations of $U(M|N)$ are given by (\ref{unum})), the double graded
commutator (\ref{N6brckrlzz}) becomes
\e\label{dbgcmtt}
[\{Q_u{}^\pu,\bar Q_\pv{}^v\},Q_w{}^\pw]=k(\d_\pv{}^\pu\d_w{}^vQ_u{}^\pw-\d_\pv{}^\pw\d_u{}^vQ_w{}^\pu).
\ee
Here we have set $Q_a=Q_u{}^\pu$, where $u=1,\cdots,M$ is a fundamental index of $U(M)$, and $\pu=1,\cdots,N$ an anti-fundamental index of $U(N)$. The structure constants can be easily read off
\e\label{strumun}
f^{ab}{}_{cd}=f_u{}^\pu{}_w{}^\pw,{}_\pv{}^v{}_\pt{}^t=k(\d_\pv{}^\pu\d_\pt{}^\pw\d_w{}^v\d_u{}^t
-\d_\pt{}^\pu\d_\pv{}^\pw\d_u{}^v\d_w{}^t).
\ee
The above structure constants were first discovered by BL \cite{Bagger08:3Alg}, by using the  following matrix realization of the hermitian 3-algebra:
\begin{eqnarray}\label{BL3Bracket}
[X, Y; \bar{Z}]_\pu{}^u=-k(X\bar{Z}Y-Y\bar{Z}X)_\pu{}^u.
\end{eqnarray}
Here $X_\pu{}^v$ is an $N\times M$ matrix, with $\bar X_v{}^\pu$ its hermitian conjugate. The right hand side of (\ref{BL3Bracket}) is the ordinary matrix multiplication. Writing $X$ as $X=X_\pu{}^vB_v{}^\pu$, with $B_v{}^\pu$ a set of basis matrices, then the 3-bracket (\ref{BL3Bracket}) is equivalent to
\e
[B_u{}^\pu,B_w{}^\pw; \bar B_\pv{}^v]=k(\d_\pv{}^\pu\d_w{}^vB_u{}^\pw-\d_\pv{}^\pw\d_u{}^vB_w{}^\pu).
\ee
Comparing the above equation with (\ref{dbgcmtt}), we see that the structure constants specified by (\ref{BL3Bracket}) are precisely the same as (\ref{strumun}).

Substituting
(\ref{strumun}) into (\ref{N6Lagrangian}) and (\ref{N6susy}) give the $\CN=6, U(M)\times U(N)$ theory \cite{Bagger08:3Alg}. In the special case of $M=N$, it becomes the well-known ABJM theory \cite{ABJM}, which has been conjectured to be the dual gauge theory
of the low energy limit of $N$ M2-branes probing a
$\textbf{C}^4/\textbf{Z}_k$ singularity. In the limit of
$N\rightarrow\infty$, the ABJM theory is dual to the M-theory on
$AdS_4\times S^7/\textbf{Z}_k$ \cite{ABJM}.

For a more mathematical approach to the relation between the superalgebra and the $\CN=6$
hermitian 3-algebra, see Ref. \cite{Jakob}.
\subsection{$\CN=8$ Theory in Terms of the Bosonic Part of $PSU(2|2)$}\label{SlgrlBLG}
In Ref. \cite{Lambert1001,Bagger10}, the Nambu 3-algebra was represented in
terms of a set of $SU(2)\times SU(2)$ $\s$-matrices. In this
section, we show explicitly that the Nambu 3-algebra can be also realized
in term of $PSU(2|2)$; and we construct the $\CN=8$ BLG theory by using several different ways.
The commutation rules of $PSU(2|2)$ are the following
\e\label{psu22}\nonumber
&&[M_{\a\b},M_{\g\d}]=\frac{1}{2}(\ep_{\a\g}M_{\b\d}+\ep_{\b\g}M_{\a\d}+\ep_{\a\d}M_{\b\g}+
\ep_{\b\d}M_{\a\g}),\nonumber\\
&&[M_{\dot\a\dot\b},M_{\dot\g\dot\d}]=\frac{1}{2}(\ep_{\dot\a\dot\g}M_{\dot\b\dot\d}
+\ep_{\dot\b\dot\g}M_{\dot\a\dot\d}+\ep_{\dot\a\dot\d}M_{\dot\b\dot\g}+
\ep_{\dot\b\dot\d}M_{\dot\a\dot\g}),\nonumber\\
&&[M_\a{}^\gamma,Q_\b{}^{\dot\b}]=\d_\b{}^\g
Q_\a{}^{\dot\b}-\frac{1}{2}\d_\a{}^\g Q_\b{}^{\dot\b},\quad [M_\a{}^\gamma,\bar Q_{\dot \b}{}^{\b}]=-\d_\a{}^\b
\bar Q_{\dot\b}{}^{\g}+\frac{1}{2}\d_\a{}^\g\bar Q_{\dot\b}{}^{\b},\nonumber\\
&&[M_{\dot\d}{}^{\dot\b},Q_\a{}^{\dot\g}]=-\d_{\dot\d}{}^{\dot\g}
Q_\a{}^{\dot\b}+\frac{1}{2}\d_{\dot\d}{}^{\dot\b} Q_\a{}^{\dot\g},\quad [M_{\dot\d}{}^{\dot\b},\bar Q_{\dot\g}{}^{\a}]=\d_{\dot\g}{}^{\dot\b}
\bar Q_{\dot\d}{}^{\a}-\frac{1}{2}\d_{\dot\d}{}^{\dot\b}\bar Q_{\dot\g}{}^{\a},\\
&&\{Q_\a{}^{\dot\a},\bar
Q_{\dot\b}{}^{\b}\}=k(\d_{\dot\b}{}^{\dot\a}M_\a{}^\b+\d_\a{}^\b
M_{\dot\b}{}^{\dot\a}),\quad \{Q_\a{}^{\dot\a},
Q_{\b}{}^{\dot\b}\}=\{\bar Q_{\dot\a}{}^{\a},
\bar Q_{\dot\b}{}^{\b}\}=0\nonumber,\ee
where $\a,\dot\b=1,2$ are $SU(2)\times SU(2)$ indices. We use the
antisymmetric matrix $\ep_{\a\b}$ ($\ep_{\dot\a\dot\b}$) to lower
undotted (dotted) indices. For example, $M_{\a\b}=\ep_{\b\g}M_\a{}^\g$ and $M_{\dot\a\dot\b}=\ep_{\dot\b\dot\g}M_{\dot\a}{}^{\dot\g}$. The inverse of $\ep_{\a\b}$ ($\ep_{\dot\a\dot\b}$) is defined as $\ep^{\a\b}$ ($\ep^{\dot\a\dot\b}$), satisfying $\ep_{\a\g}\ep^{\g\b}=\d_\a{}^\b$ ($\ep_{\dot\a\dot\g}\ep^{\dot\g\dot\b}=\d_{\dot\a}{}^{\dot\b}$). Note that the superalgebra (\ref{psu22}) takes the form of (\ref{SLie6}).
Since we wish to explicitly construct the Nambu 3-algebra in terms of $PSU(2|2)$, it is
useful to define
\begin{equation}\label{imbedding}
Q^a=\frac{1}{2}\s^{a\dag}{}_{\dot\a}{}^\a Q_\a{}^{\dot\a},\quad \bar
Q^a=\frac{1}{2}\s^{a}{}_{\a}{}^{\dot\a}\bar Q_{\dot\a}{}^{\a},\quad M^{ab}=-(\s^{ab}{}_\a{}^\b M_\b{}^\a+\bar\s^{ab}{}_{\dot\a}{}^{\dot\b}M_{\dot\b}{}^{\dot\a}).
\end{equation}
Here the $SU(2)\times SU(2)$
$\s$-matrices are given by
\e &&\nonumber\sigma^a{}_\a{}^{\dot\a}=(\sigma^1,\sigma^2,\sigma^3,
i\mathbb{I}),\quad\quad\quad\quad\quad
\sigma^{a\dag}{}_{\dot\a}{}^{\a}=(\sigma^1,\sigma^2,\sigma^3,
-i\mathbb{I})\\
&&\s^{ab}{}_{\a}{}^\b=\frac{1}{4}(\s^a\s^{b\dag}-\s^b\s^{a\dag})_{\a}{}^\b,\quad\quad
\bar\s^{ab}{}_{\dot\a}{}^{\dot\b}=\frac{1}{4}(\s^{a\dag}\s^{b}-\s^{b\dag}\s^{a})_{\dot\a}{}^{\dot\b},\ee
where $\s^{ab}$ and $\bar\s^{ab}$ satisfy the further `duality'
conditions
\begin{equation}
\s^{ab}=-\frac{1}{2}\vp^{abcd}\s_{cd},\quad\quad
\bar\s^{ab}=\frac{1}{2}\vp^{abcd}\bar\s_{cd}.
\end{equation}
Our convention is that $\vp^{1234}=1$ and $\vp_{abcd}=\vp^{abcd}$. With these definitions, we are able to recast the superalgebra (\ref{psu22}) into the form
\e\label{psu22n}
&&[M^{ab},M^{cd}]=\d^{bc}M^{ad}-\d^{ac}M^{bd}-\d^{bd}M^{ac}+\d^{ad}M^{bc},\nonumber\\
&&[M^{ab},Q^c]=\d^{bc}Q^a-\d^{ac}Q^b,\quad [M^{ab},\bar Q^c]=\d^{bc}\bar Q^a-\d^{ac}\bar Q^b,\nonumber\\
&&\{Q^a,\bar Q^b\}=-\frac{k}{4}\vp^{abcd}M_{cd},\quad \{Q^a,Q^b \}=\{\bar Q^a,\bar Q^b\}=0.
\ee
Here
\begin{equation}\label{dlt}
M_{cd}\equiv \d_{ca}\d_{db}M^{ab}.
\end{equation}
It is still necessarily to check the Jacobi identities of (\ref{psu22n}). The $MMM$ Jacobi identity is obviously satisfied. It is straightforward to verify that the $MMQ$ ($MM\bar Q$) Jacobi identity is obeyed. The total antisymmetry of $\vp^{abcd}$ guarantees that the $QQ\bar Q$ $(\bar Q\bar QQ)$ Jacobi identity is obeyed. To check the $MQ\bar Q$ Jacobi identity, we define
\e
M^{ab}_-=-\s^{ab}{}_\a{}^\b M_\b{}^\a,\quad M^{ab}_+=-\bar\s^{ab}{}_{\dot\a}{}^{\dot\b}M_{\dot\b}{}^{\dot\a}.
\ee
With these definitions, we note that
\begin{equation}\label{mpm}
M^{ab}_{\pm}=\pm \frac{1}{2}\vp^{abcd}M^{cd}_\pm,\quad[M^{ab}_+,M^{ab}_-]=0,\quad M^{ab}=M^{ab}_+ + M^{ab}_-,\quad \frac{1}{2}\vp^{abcd}M_{cd}=M^{ab}_+ - M^{ab}_-.
\end{equation}
Using the above equations, it is not difficult to prove that the $MMQ$ ($MM\bar Q$) Jacobi identity is obeyed. Now every Jacobi identity of (\ref{psu22n}) is satisfied. Therefore the superalgebra (\ref{psu22n}) is closed.

A short calculation gives
\begin{eqnarray}\label{Nmbbrck}
&&[t^b, t^c; \bar t^a]\doteq[\{Q^b,\bar Q^a\},
Q^c]=-\frac{k}{2}\vp^{bcad}Q_d\doteq-\frac{k}{2}\vp^{bcad}t_d.
\end{eqnarray}
Namely, the double graded commutator is indeed a realization of the
Nambu 3-bracket. In this realization, the totally antisymmetric structure constants are given by
\begin{equation}\label{N8strt}
f^{abcd}=-\frac{k}{2}\vp^{abcd}.
\end{equation}
Also, the FI satisfied by the Nambu 3-bracket is
equivalent to the $MMQ$ Jacobi identity of (\ref{psu22n}), as we
proved in section \ref{Slgrlzhm3}. Therefore the Nambu 3-algebra is
realized in terms of the superalgebra $PSU(2|2)$. Hence the
bosonic part of $PSU(2|2)$, $SO(4)\cong SU(2)\times SU(2)$, is the
Lie algebra of the gauge group of the $\CN=8$ BLG theory. And the
matter fields are in the vector representation of $SO(4)$. Substituting (\ref{N8strt}) into (\ref{N6Lagrangian}) and (\ref{N6susy}) gives the $\CN=8$ BLG theory. (In this paper, we re-scale $A_\mu{}^a{}_b$ by a factor $\frac{1}{k}$, i.e., $A_\mu{}^a{}_b\rightarrow \frac{1}{k}A_\mu{}^a{}_b$.) The
same theory is obtained in Ref. \cite{HosomichiJD} by promoting the
$\CN=4$ supersymmetry to $\CN=8$. The $\CN=8$ BLG theory was assumed to be the
dual gauge description of two M2-branes.

Eq. (\ref{Nmbbrck}) may be counterintuitive at first sigh, since the
anticommutator satisfies $\{Q^b,\bar Q^a\}=\{\bar Q^a,Q^b\}$, i.e.,
it seems that it is symmetric in $ab$. However, there is no clash
with fact that (\ref{Nmbbrck}) is antisymmetric in $ab$ if we
notice that
\begin{eqnarray}
\{Q^b,\bar Q^a\}=-\frac{k}{4}\vp^{bacd}M_{cd}=-\{Q^a,\bar
Q^b\}=\{\bar Q^a,Q^b\},
\end{eqnarray}
namely, the last two anticommutators are \emph{different}.

One can of course calculate the double graded commutator directly by using (\ref{psu22}):
\e
[\{Q_\a{}^{\dot\a},\bar
Q_{\dot\g}{}^{\g}\}, Q_\b{}^{\dot\b}]=k(\d_{\dot\g}{}^{\dot\a}\d_\b{}^\g Q_\a{}^{\dot\b}
-\d_\a{}^\g \d_{\dot\g}{}^{\dot\b}Q_\b{}^{\dot\a}).
\ee
Here the structure constants are given by
\e\label{N8strt4}
f_\a{}^{\dot\a}{}_\b{}^{\dot\b},{}_{\dot\g}{}^\g{}_{\dot\d}{}^\d
=k(\d_\a{}^\d\d_\b{}^\g\d_{\dot\g}{}^{\dot\a}\d_{\dot\d}{}^{\dot\b}
-\d_\a{}^\g\d_\b{}^\d\d_{\dot\g}{}^{\dot\b}\d_{\dot\d}{}^{\dot\a}).
\ee
Substituting the above structure constants into (\ref{N6Lagrangian}) and (\ref{N6susy}) also gives the $\CN=8$ BLG theory. In this formalism, the matter fields are in the bi-fundamental representation of $SU(2)\times SU(2)$. We can convert (\ref{N8strt4}) into (\ref{N8strt}) by using the equation
\e
f_\a{}^{\dot\a}{}_\b{}^{\dot\b},{}_{\dot\g}{}^\g{}_{\dot\d}{}^\d
(\frac{1}{2}\s^{a\dag}{}_{\dot\a}{}^\a)(\frac{1}{2}\s^{b\dag}{}_{\dot\b}{}^\b)
(\frac{1}{2}\s^c{}_\g{}^{\dot\g})(\frac{1}{2}\s^d{}_\d{}^{\dot\d})=f^{abcd}.
\ee
The RHS is given by
\e
\frac{k}{8}[{\rm Tr}(\s^{a\dag}\s^d\s^{b\dag}\s^c-\s^{a\dag}\s^c\s^{b\dag}\s^d)]=-\frac{k}{2}\vp^{abcd},
\ee
which is in agreement with (\ref{N8strt}).

One can also derive the $\CN=8$ BLG theory in an alternatively way, by reading off the representation matrices $(\t^{ab})^{cd}$, invariant quadratic form $k_{ab,cd}$ and structure constants $C^{ab,cd}{}_{ef}$ from (\ref{psu22n}), and substituting them into the Lagrangian (\ref{N6LagrangianLie}) and the SUSY transformations (\ref{N6susyLie}). The representation matrices $(\t^{ab})^{cd}$ determined by the fermionic generators (see the second line of (\ref{psu22n})) are given by
\begin{equation}\label{untld}
(\t^{ab})^{cd}=\d^{ac}\d^{bd}-\d^{bc}\d^{ad}.
\end{equation}
Comparing $\{Q^a,\bar Q^b\}=(\t^{cd})^{ab}k_{cd,ef}M^{ef}$ with the first equation of the third line of (\ref{psu22n}), we obtain the invariant quadratic form
\e\label{qdtcfm}
k_{cd,ef}=-\frac{k}{8}\vp_{cdef}=\frac{1}{4}f_{cdef}.
\ee
So the invariant quadratic form is nothing but the structure constants of the Nambu 3-algebra up to a constant. Since $k_{ab,cd}$ is an invariant form on the superalgebra (\ref{psu22n}), we can define
\begin{equation}
\tilde M_{ab}\equiv k_{ab,cd}M^{cd}.
\end{equation}
Namely, if we use $k_{ab,cd}$ to lower the indices of a generator, we
put a tilde on the generator. However, if we use $\d_{ab}$ to lower the indices of a generator, we do \emph{not} put a tilde on the generator (see (\ref{dlt})). Note that
\e\label{tld}
(\tilde\t^{ab})^{cd}\equiv \d^{ae}\d^{df}k_{ef,gh}(\t^{gh})^{cd}=-\frac{k}{4}\vp^{abcd}.
\ee
are quite different from that of (\ref{untld}). So the quadratic form $k_{ab,cd}$ plays a fundamental role in constructing the theory.

The structure constants $C^{ab,cd}{}_{ef}$ can be read off from the first line of (\ref{psu22n}). However, we actually used $\tilde C_{ab,cd,ef}\equiv k_{ab,\pa\pb}k_{cd,\pc\pd}C^{\pa\pb,\pc\pd}{}_{ef}$ to construct the theory (see Eq. (\ref{tdC})). After some work, we obtain
\e\label{tdCBLG}
\tilde C_{ab,cd,ef}&=&\frac{k^2}{32}(\vp_{abe}{}^g\vp_{cdgf}-\vp_{abf}{}^g\vp_{cdge})\\
&=&\frac{1}{8}(f_{abe}{}^gf_{cdgf}-f_{abf}{}^gf_{cdge})\nonumber.
\ee
The gauge field $\hat A^m_\mu$ defined by (\ref{gauge}) is given by
\e\label{hatgauge}
\hat A^{ab}_\mu=A_\mu{}^c{}_d(\t^{ab})^d{}_c=-2A_\mu^{[ab]}.
\ee
Substituting (\ref{untld}), (\ref{qdtcfm}), (\ref{tdCBLG}) and (\ref{hatgauge}) into the Lagrangian (\ref{N6LagrangianLie}) and the SUSY transformations (\ref{N6susyLie}) reproduces the $\CN=8$ BLG theory.

As a consistent check, one can also calculate the structure constants of the Nambu 3-algebra by using Eq. (\ref{sltN6str}):
\e\label{N8strt2}
f^{abcd}=k_{ef,gh}(\t^{ef})^{ad}(\t^{gh})^{bc}=-\frac{k}{2}\vp^{abcd},
\ee
which are indeed the same as (\ref{N8strt}).

With (\ref{mpm}), we can recast the superalgebra (\ref{psu22n}) into the following form
\e\label{psu22nn}
&&[M_\pm^{ab},M_\pm^{cd}]=\d^{bc}M_\pm^{ad}-\d^{ac}M_\pm^{bd}-\d^{bd}M_\pm^{ac}
+\d^{ad}M_\pm^{bc},\quad [M^{ab}_+,M^{cd}_-]=0,\nonumber\\
&&[M^{ab}_\pm,Q^c]=\frac{1}{2}(\d^{bc}\d^{ad}-\d^{ac}\d^{bd}\mp\vp^{abcd})Q_d,
\quad [M^{ab}_\pm,\bar Q^c]=\frac{1}{2}(\d^{bc}\d^{ad}-\d^{ac}\d^{bd}\mp\vp^{abcd})\bar Q_d,\nonumber\\
&&\{Q^a,\bar Q^b\}=-\frac{k}{2}(M^{ab}_+ -M^{ab}_-),\quad \{Q^a,Q^b \}=\{\bar Q^a,\bar Q^b\}=0.
\ee
Using $M^{ab}_\pm=\frac{1}{2}(M^{ab}\pm\frac{1}{2}\vp^{abcd}M^{cd})$, it is not difficult to prove that every Jacobi identity is obeyed. Let us now try to read off the representation matrices $(\t^{ab}_\pm)^{cd}$, invariant quadratic forms $k_{\pm ab,cd}$ and structure constants $C_{\pm}^{ab,cd}{}_{ef}$ from (\ref{psu22nn}).
The representation matrices can be easily read off from the second line of (\ref{psu22nn}).
\e\label{tpm}
(\t^{ab}_\pm)^{cd}=\frac{1}{2}(\d^{ac}\d^{bd}-\d^{bc}\d^{ad}\pm\vp^{abcd}).
\ee
To determine the quadratic forms $k_{\pm ab,cd}$, we re-write the first anti-commutator of the third line of (\ref{psu22nn}) as
\e
\{Q^a,\bar Q^b\}=(\t^{cd}_+)^{ab}k_{+cd,ef}M^{ef}_+ + (\t^{cd}_-)^{ab}k_{-cd,ef}M^{ef}_-.
\ee
Because of the duality conditions
\e\label{duatao}
(\t^{cd}_{\pm})^{ab}=\pm \frac{1}{2}\vp^{cdgh}(\t^{gh}_\pm)^{ab},\quad M^{ef}_{\pm}=\pm \frac{1}{2}\vp^{efgh}M^{gh}_\pm,
\ee
we must impose the duality conditions on the quadratic forms
\e
k_{\pm gh,ef}=\pm\frac{1}{2}\vp_{ghcd}k_{\pm cd,ef},\quad k_{\pm cd,gh}=\pm\frac{1}{2}\vp_{ghef}k_{\pm cd,ef},
\ee
which lead us to the solutions:
\e\label{kpm}
k_{\pm ab,cd}=\mp\frac{k}{16}(\d^{ac}\d^{bd}-\d^{bc}\d^{ad}\pm\vp^{abcd}).
\ee
Since the bosonic generators $M^{ab}_{\pm}$ satisfy the duality conditions
$
M^{ab}_{\pm}=\pm\frac{1}{2}\vp^{abcd}M^{cd}_{\pm},
$
we must impose the duality conditions on the structure constants:
\e\label{duaC}
C_{\pm}^{ab,cd}{}_{ef}=\pm\frac{1}{2}\vp^{abgh}C_{\pm}^{gh,cd}{}_{ef}
=\pm\frac{1}{2}\vp^{cdgh}
C_{\pm}^{ab,gh}{}_{ef}=\pm\frac{1}{2}
\vp_{efgh}C_{\pm}^{ab,cd}{}_{gh}.
\ee
The structure constants of the bosonic subalgebra of (\ref{psu22nn}) can be read off from the $MMQ$ Jacobi identities of (\ref{psu22nn}):
\e
[[M^{ab}_\pm,M^{cd}_\pm],Q^e]+[[M^{cd}_\pm,Q^e],M^{ab}_\pm]+[[Q^e,M^{ab}_\pm],M^{cd}_\pm]=0.
\ee
A short calculation gives
\e
C_{\pm}^{ab,cd}{}_{gh}(\t_\pm^{gh})^{ef}=[\t^{ab}_\pm,\t^{cd}_\pm]^{ef}.
\ee
Substituting $(\ref{tpm})$ into the LHS and using the duality conditions $(\ref{duaC})$, the above equation becomes
\e
2C_{\pm}^{ab,cd}{}_{ef}=[\t^{ab}_\pm,\t^{cd}_\pm]_{ef}.
\ee
Multiplying both sides by $(\t^{gh}_\pm)^{ef}$, the structure constants can be recast into
\e
C_{\pm}^{ab,cd}{}_{gh}=\frac{1}{4}[\t^{ab}_\pm,\t^{cd}_\pm]_{ef}(\t^{gh}_\pm)^{ef}=
-\frac{1}{4}{\rm Tr}([\t^{ab}_\pm,\t^{cd}_\pm]\t_{\pm gh}).
\ee
Here $\t_{\pm gh}\equiv \d_{ge}\d_{hf}\t_{\pm}^{ef}$. With the first two equations of $(\ref{duatao})$, the structure constants satisfy the duality conditions (\ref{duaC}) manifestly. It is straightforward to verify that
\e
-\frac{1}{4}{\rm Tr}([\t^{ab}_\pm,\t^{cd}_\pm]\t_{\pm gh})M^{gh}_\pm=\d^{bc}M_\pm^{ad}-\d^{ac}M_\pm^{bd}-\d^{bd}M_\pm^{ac}
+\d^{ad}M_\pm^{bc},
\ee
which are exactly the same as the right-hand sides of the first two equation of (\ref{psu22nn}). After some algebraic steps, we obtain
\e\label{strun8}
\tilde C_{\pm ab,cd,ef}&=&k_{\pm ab,\pa\pb}k_{\pm cd,\pc\pd}C_{\pm}^{\pa\pb,\pc\pd}{}_{ef}\nonumber\\&=&
-\frac{k^2}{64}{\rm Tr}([\t_{\pm ab},\t_{\pm cd}]\t_{\pm ef})\\ \nonumber
&=&\frac{k^2}{64}[-\d_{ae}(\t_{\pm cd})_{bf}+\d_{af}(\t_{\pm cd})_{be}+\d_{be}(\t_{\pm cd})_{af}-\d_{bf}(\t_{\pm cd})_{ae}\\ \nonumber&&\quad\quad+\d_{ce}(\t_{\pm ad})_{df}-\d_{cf}(\t_{\pm ab})_{de}+\d_{de}(\t_{\pm ab})_{cf}-\d_{df}(\t_{\pm ab})_{ce}],
\ee
where $(\t_{\pm cd})_{bf}=\frac{1}{2}(\d_{cb}\d_{df}-\d_{cf}\d_{db}\pm \vp_{cdbf})$.
The gauge fields $\hat A^m_{\mu\pm}$ defined by (\ref{gauge}) are given by
\e\label{Amp}
\hat A^{ab}_{\mu\pm}=A_\mu{}^c{}_d(\t^{ab}_\pm)^d{}_c=-(A_\mu^{[ab]}\pm\frac{1}{2}\vp^{abcd}A^{cd}_\mu)
=\pm\frac{1}{2}\vp^{abcd}\hat A^{cd}_{\mu\pm}.
\ee
Substituting (\ref{tpm}), (\ref{kpm}), (\ref{strun8}) and (\ref{Amp}) into the Lagrangian (\ref{N6LagrangianLie}) and the SUSY transformations (\ref{N6susyLie}) also reproduces the $\CN=8$ BLG theory.

As another consistent check, we may calculate the structure constants of the Nambu 3-algebra by substituting (\ref{tpm}) and (\ref{kpm}) into Eq. (\ref{sltN6str}):
\e\label{N8strt3}
f^{abcd}=k_{+ef,gh}(\t_+^{ef})^{ad}(\t_+^{gh})^{bc}+k_{-ef,gh}(\t_-^{ef})^{ad}(\t_-^{gh})^{bc}
=-\frac{k}{2}\vp^{abcd}.
\ee
They are precisely the same as (\ref{N8strt}) or (\ref{N8strt2}).

Finally, if we define
\e
N^{ab}=\frac{1}{2}\vp^{abcd}M_{cd},\ee
then the superalgebra (\ref{psu22n}) can be recast into
\e\label{psu22n4}
&&[N^{ab},N^{cd}]=\vp^{abd}{}_eN^{ec}-\vp^{abc}{}_eN^{ed},\nonumber\\
&&[N^{ab},Q^c]=-\vp^{abcd}Q_d,\quad [N^{ab},\bar Q^c]=-\vp^{abcd}\bar Q_d ,\nonumber\\
&&\{Q^a,\bar Q^b\}=-\frac{k}{2}N^{ab},\quad \{Q^a,Q^b \}=\{\bar Q^a,\bar Q^b\}=0.
\ee
One can verify that every Jacobi identity is obeyed. The representation matrices $(\t^{ab})^{cd}$, invariant quadratic form $k_{ab,cd}$ and structure constants $\tilde C_{ab,cd,ef}$, read off from (\ref{psu22n4}), are give by
\e\label{tkC}
&&(\t^{ab})^{cd}=\vp^{abcd},\quad k_{ab,cd}=-\frac{k}{8}\vp_{abcd},\\ \nonumber
&&\tilde C_{ab,cd,ef}=\frac{k^2}{32}(\d_{ae}\vp_{bcdf}-\d_{be}\vp_{acdf}-\d_{af}\vp_{bcde}
+\d_{bf}\vp_{acde}).
\ee
One unusual feature is that both $(\t^{ab})^{cd}$ and $k^{ab,cd}$ are proportional to $\vp^{abcd}$. However, by the definition $\tilde N_{ab}=k_{ab,cd}N^{cd}$, we learn that
\begin{equation}\label{tldtao}
(\tilde \t^{ab})^{cd}=-\frac{k}{4}(\d^{ac}\d^{bd}-\d^{bc}\d^{ad}).
\end{equation}
Comparing (\ref{tldtao}) and the first equation of (\ref{tkC}) with (\ref{untld}) and (\ref{tld}), respectively, we note that the roles of $\t^{ab}$ and $\tilde \t^{ab}$ are switched; this due to the definitions $N^{ab}=\frac{1}{2}\vp^{abcd}M_{cd}$ and $\tilde N_{ab}=k_{ab,cd}N^{cd}=-\frac{k}{4}M_{ab}$.

The gauge fields $\hat A^m_{\mu}$ defined by (\ref{gauge}) are given by
\e\label{Amp22}
\hat A^{ab}_{\mu}=A_\mu{}^c{}_d(\t^{ab})^d{}_c=-\vp^{abcd}A^{cd}_\mu.
\ee
Substituting (\ref{tkC}) and (\ref{Amp22}) into the Lagrangian (\ref{N6LagrangianLie}) and the SUSY transformations (\ref{N6susyLie}) also reproduces the $\CN=8$ BLG theory.

As the final consistent check, we calculate the structure constants of the Nambu 3-algebra by substituting the first two equations of (\ref{tkC}) into Eq. (\ref{sltN6str}):
\e\label{N8strt4}
f^{abcd}=k_{ef,gh}(\t^{ef})^{ad}(\t^{gh})^{bc}
=-\frac{k}{2}\vp^{abcd}.
\ee
They are precisely the same as (\ref{N8strt}), (\ref{N8strt2}) or (\ref{N8strt3}).

\section{A Quantization Scheme for the 3-brackets}\label{quantize}
It is well known that a 3-bracket such as Nambu 3-bracket is difficult to
quantize. However, since a 3-algebra can be constructed in terms of a superalgebra, we may quantize a 3-algebra system by quantizing the corresponding superalgebra system. For example, if we promote the fermionic and bosonic generators of (\ref{psu22n}) as quantum mechanical operators, and promote (\ref{Nmbbrck}) as a quantum
mechanical double graded commutator, our approach may provide a
quantization scheme for the Nambu 3-bracket. Similarly, the
3-brackets of the symplectic and hermitian 3-algebras may be
quantized in the same fashion.

\section{Conclusions}\label{conclusions}
In this paper, we used superalgebras to realize the 3-algebras used to construct ${\cal N}=6, 8$ CSM theories. We first worked out the general commutation relations of these superalgebras by decomposing the superalgebra used to realize the 3-algebra in $\CN=5$ theory. Using this superalgebra realization of 3-algebras, i.e., $t^b\doteq Q^b$ $[t^b, t^c; \bar t_a]\doteq[\{Q^b,\bar Q_a\},
Q^c]$, we were able to rederive the general ${\cal N}=6$ CSM theory and the $\CN=8$ BLG theory in terms of ordinary Lie algebras from their 3-algebra counterparts. In this realization, the Lie algebras of the Lie groups are nothing but the bosonic parts of the superalgebras, and the representations are determined by the fermionic generators. Specifically, we used $U(M|N)$ and $OSp(2|2N)$ to realize the $\CN=6$ Hermitian 3-algebra, and used $PSU(2|2)$ to represent the Nambu 3-algebra whose structure constants are totally antisymmetric. The Nambu 3-bracket is constructed explicitly in terms of a double graded commutator of $PSU(2|2)$. The $\CN=8$ BLG theory with $SO(4)$ gauge group was constructed by using several different ways. In summary, we have demonstrated that the superalgebra realization of the 3-algebras provides a unified framework for classifying the gauge groups of the $\CN=5, 6, 8$ theories based on 3-algebras.

We also proposed a quantization scheme for the 3-brackets, by promoting the fermionic generators $Q^a$ and $\bar Q_b$ as quantum mechanical operators, and by promoting the double graded commutator $[\{Q^b,\bar Q_a\},
Q^c]$ as quantum mechanical double graded commutator.

It would be interesting to investigate the relation between the superalgebra and 3-algebra further, and explore the physical meaning. It would be nice to investigate the relation between the matrix realization of 3-algebra and the superalgebra realization of 3-algebra, since the superalgebras may have matrix representations.

\section{Acknowledgement} We would like to thank Yong-Shi Wu for
useful discussions.

\appendix

\section{Conventions and Useful Identities}\label{Identities}
The conventions and useful identities are adopted from our previous
paper \cite{ChenWu3}. In $1+2$ dimensions, the gamma matrices are defined as
\begin{equation}
(\gamma_{\mu})_{\alpha}{}^\gamma(\gamma_{\nu})_{\gamma}{}^\beta+
(\gamma_{\nu})_{\alpha}{}^\gamma(\gamma_{\mu})_{\gamma}{}^\beta=
2\eta_{\mu\nu}\delta_{\alpha}{}^\beta.
\end{equation} For the metric we
use the $(-,+,+)$ convention. The gamma matrices in the Majorana
representation can be defined in terms of Pauli matrices:
$(\gamma_{\mu})_{\alpha}{}^\beta=(i\sigma_2, \sigma_1, \sigma_3)$,
satisfying the important identity
\begin{equation}
(\gamma_{\mu})_{\alpha}{}^\gamma(\gamma_{\nu})_{\gamma}{}^\beta
=\eta_{\mu\nu}\delta_{\alpha}{}^\beta+\varepsilon_{\mu\nu\lambda}(\gamma^{\lambda})_{\alpha}{}^\beta.
\end{equation}
We also define
$\varepsilon^{\mu\nu\lambda}=-\varepsilon_{\mu\nu\lambda}$. So
$\varepsilon_{\mu\nu\lambda}\varepsilon^{\rho\nu\lambda} =
-2\delta_\mu{}^\rho$. We raise and lower spinor indices with an
antisymmetric matrix
$\epsilon_{\alpha\beta}=-\epsilon^{\alpha\beta}$, with
$\epsilon_{12}=-1$. For example,
$\psi^\alpha=\epsilon^{\alpha\beta}\psi_\beta$ and
$\gamma^\mu_{\alpha\beta}=\epsilon_{\beta\gamma}(\gamma^\mu)_\alpha{}^\gamma
$, where $\psi_\beta$ is a Majorana spinor. Notice that
$\gamma^\mu_{\alpha\beta}=(\mathbb{I}, -\sigma^3, \sigma^1)$ are
symmetric in $\alpha\beta$. A vector can be represented by a
symmetric bispinor and vice versa:
\begin{equation}
A_{\alpha\beta}=A_\mu\gamma^\mu_{\alpha\beta},\quad\quad A_\mu=-\frac{1}{2}\gamma^{\alpha\beta}_\mu A_{\alpha\beta}.
\end{equation}
We use the following spinor summation convention:
\begin{equation}
\psi\chi=\psi^\alpha\chi_\alpha,\quad\quad
\psi\gamma_\mu\chi=\psi^\alpha(\gamma_{\mu})_{\alpha}{}^\beta\chi_\beta,
\end{equation}
where $\psi$ and $\chi$ are anticommuting Majorana spinors. In
$1+2$ dimensions the Fierz transformation reads
\begin{eqnarray}
(\lambda\chi)\psi &=& -\frac{1}{2}(\lambda\psi)\chi -\frac{1}{2}
(\lambda\gamma_\nu\psi)\gamma^\nu\chi.
\end{eqnarray}
\section{The Commutation Relations of $OSp(2|2N)$ and $U(M|N)$}\label{superalgebras}
\subsection{$OSp(2|2N)$}\label{cmrsp22n}
The commutation relations of $OSp(2|2N)$ are given by
\begin{eqnarray}\label{sp22n}
&&[M_{ij},M_{kl}]=0,\\
&&[M_{ab},M_{cd}]=\om_{bc}M_{ad}+\om_{ac}M_{bd}+\om_{ad}M_{bc}+\om_{bd}M_{ac},\nonumber\\
&&[M_{ij},Q_{ck}]=\d_{jk}Q_{ci}-\d_{ik}Q_{cj},\nonumber\\
&&[M_{ab},Q_{ck}]=\om_{ac}Q_{bk}+\om_{bc}Q_{ak},\nonumber\\
&&\{Q_{ai},Q_{bj}\}=k(\om_{ab}M_{ij}+\d_{ij}M_{ab}).
\end{eqnarray}
Here $a=1,\cdots,2N$ is an $Sp(2N)$ index, and $i=1,2$ an $SO(2)$ index. And $\omega_{ab}$ is the invariant antisymmetric tensor of $Sp(2N)$.  With $\om_{ai,bj}=\om_{ab}\d_{ij}=\om_{IJ}$, the superalgebra (\ref{sp22n}) takes the form of (\ref{SLie26}). To convert (\ref{sp22n}) into the form of (\ref{SLie6}), we by combine the fermionic generators as follows:
\e
Q_{a\pm}=\frac{1}{\sqrt{2}}(Q_{a1}\pm iQ_{a2}).
\ee
Now (\ref{sp22n}) becomes
\begin{eqnarray}\label{sp22n2}
&&[M_{+-},M_{+-}]=0, \nonumber\\
&&[M_{ab},M_{cd}]=\om_{bc}M_{ad}+\om_{ac}M_{bd}+\om_{ad}M_{bc}+\om_{bd}M_{ac},\nonumber\\
&&[M_{+-},Q_{c\pm}]=\mp i\epsilon_{+-}Q_{c\pm},\\
&&[M_{ab},Q_{c\pm}]=\om_{ac}Q_{b\pm}+\om_{bc}Q_{a\pm},\nonumber\\
&&\{Q_{a+},Q_{b-}\}=k(\om_{ab}M_{+-}+h_{+-}M_{ab}),\nonumber\\
&&
\{Q_{a+},Q_{b+}\}=\{Q_{a-},Q_{b-}\}=0\nonumber,
\end{eqnarray}
where $h_{+-}=h_{-+}=1$,
$\epsilon_{+-}=-\epsilon_{-+}=ih_{+-}$, and $M_{+-}\equiv -iM_{12}$. And the invariant anti-symmetric tensor becomes $\om_{a+,b-}\equiv\om_{ab}h_{+-}$. Now the superalgebra (\ref{sp22n2}) indeed takes the form of (\ref{SLie6}).
\subsection{$U(M|N)$}\label{cmrunum}
The commutation relations of $U(M|N)$ are given by
\e\label{unum}\nonumber
&&[M_u{}^{v},M_w{}^t]=\d_w{}^vM_u{}^t-\d_u{}^tM_{w}{}^v,\quad [M_\pu{}^{\pv},M_\pw{}^\pt]=\d_\pw{}^\pv M_\pu{}^\pt-\d_\pu{}^\pt M_{\pw}{}^\pv\nonumber\\
&&[M_u{}^{v},Q_w{}^\pw]=\d_w{}^vQ_u{}^\pw,\quad [M_u{}^{v},\bar Q_\pw{}^w]=-\d_u{}^w\bar Q_\pw{}^v,\nonumber\\
&&[M_\pu{}^{\pv},Q_w{}^\pw]=-\d_\pu{}^\pw Q_w{}^\pv,\quad [M_\pu{}^{\pv},\bar Q_\pw{}^w]=\d_\pw{}^\pv \bar Q_\pu{}^w\nonumber\\
&& \{Q_u{}^\pu,\bar Q_\pv{}^v\}=k(\d_\pv{}^\pu M_u{}^v+\d_u{}^vM_{\pv}{}^\pu).
\ee
Here $Q^a=Q_u{}^\pu$, where $u=1,\cdots,M$ is a fundamental index of $U(M)$, and $\pu=1,\cdots,N$ an anti-fundamental index of $U(N)$. And the anti-symmetric tensor $\om_{IJ}$ reads
\begin{equation}
\omega_{IJ}=\begin{pmatrix} 0 & \d_u{}^v\d_{\pu}{}^\pv \\
-\d^u{}_v\d^{\pu}{}_\pv & 0
\end{pmatrix}.
\end{equation}
The superalgebra (\ref{unum}) also takes the form of (\ref{SLie6}) decomposed from (\ref{SLie26}).


\end{document}